\documentclass[a4paper,11pt]{article}
\pdfoutput=1 % if your are submitting a pdflatex (i.e. if you have
\usepackage{jinstpub}

\title{\boldmath Observation of primary scintillations in the visible range in liquid argon doped with methane}

\author[a,b]{A.~Bondar,}
\author[a,b]{E.~Borisova,}
\author[a,b]{A.~Buzulutskov,}
\author[a,b]{E.~Frolov,}
\author[a,b]{V.~Nosov,}
\author[a,b,1]{V.~Oleynikov,\note{Corresponding author.}}
\author[a,b]{A.~Sokolov}

\affiliation[a]{Budker Institute of Nuclear Physics, Lavrentiev ave. 11, Novosibirsk 630090, Russia}
\affiliation[b]{Novosibirsk State University, Pirogova st. 2, Novosibirsk 630090, Russia}

\emailAdd{A.F.Buzulutskov@inp.nsk.su}
\emailAdd{V.P.Oleynikov@inp.nsk.su}

\abstract{
	Neutron veto detector based on liquid scintillator containing hydrogen atoms is an integral part of any underground experiment for dark matter search.  So far, a flammable mixture of liquid hydrocarbons was used as a liquid scintillator in such detectors. A safe alternative might be a liquid scintillator based on liquid argon doped with methane. In this work, we have studied the primary scintillations in pure liquid argon and its mixtures with methane, the CH$_{4}$ content varying from 100 ppm to 5\%. The primary scintillations have for the first time been observed in liquid argon doped with methane, in the visible and near infrared range, and their relative light yields have been measured as a function of the CH$_{4}$ content. 
}


\begin{document}
	\maketitle
	\flushbottom
	\newcommand*{\doi}[1]{\href{http://dx.doi.org/#1}{doi: #1}}

%\linenumbers

\section{Introduction}\label{Introduction}

Detectors for direct dark matter search based on liquid noble-gases are in fact those of nuclear recoils that would be produced by collisions with hypothetical dark matter particles (WIMPs): see for example~\cite{Aalseth2018}. Since the nuclear recoils could also be produced by neutrons from environmental background, it is important to have an effective neutron veto detector containing hydrogen atoms, the latter acting as neutron moderator. So far, neutron veto detectors based on flammable mixtures of liquid organic scintillators have been used~\cite{Agnes2015,Agnes2016}.  

A safe alternative might be a liquid scintillator based on liquid argon doped with methane. This idea was discussed for application in the DarkSide experiment~\cite{Galbiati_Private_communication}. In particular, according to estimates, the CH$_4$ content of 5-10\% in liquid Ar would be enough for compact neutron veto detector of about 1~m thick.

However, it is well known that even tiny amount ($>$0.1~ppm) of CH$_4$ dopant would immediately quench primary scintillations in liquid Ar~\cite{Jones2013} in the  vacuum ultraviolet (VUV) range, i.e. those due to ``standard'' mechanism provided by excimer (Ar$_2^*$) photon emission~\cite{Chepel2013,Buzulutskov2017}. The quenching is due to both VUV light absorption on CH$_4$ molecules~\cite{Jones2013,SpectralAtlas} (main mechanism) and excimer deexcitation in collisions with CH$_4$.

On the other hand, some authors observed unusual primary scintillations in liquid Ar, namely weak scintillations in the visible and near infrared (NIR) range \cite{Heindl2011,Buzulutskov2011,StudyInfraredScintillations2012P2,Alexander2016}. Their nature is yet unclear. In~\cite{Buzulutskov2018} it was suggested that such primary scintillations of liquid Ar in the non-VUV might be explained by neutral bremsstrahlung (NBrS) emission of primary ionization electrons, decelerated in the medium down to the energy domain of the NBrS effect ($\sim$10~eV). 

So far, the NBrS emission in the form of electroluminescence under moderate and high electric fields was theoretically studied in~\cite{Buzulutskov2018}, with predicted continuous spectrum in the visible and NIR range, and experimentally observed in gaseous Ar in~\cite{Buzulutskov2018,Bondar2020,Tanaka2020}. 

In addition, the NBrS radiation induced by primary ionization electrons was supposed to be present not only in atomic noble gases, but in molecular gases as well, namely in air~\cite{Samarai2016}. Similarly, we may suppose that the NBrS emission could also exist in a liquid mixture of Ar and CH$_4$ in the form of primary scintillations in the visible and NIR range. 

The purpose of the present work is to test this hypothesis: we studied the primary scintillations, in the visible and NIR range, in pure liquid Ar and its mixtures with CH$_4$, the CH$_{4}$ content varying from 100 ppm to 5\%. 

\section{Experimental setup}

\begin{figure}[!htb]
	\center{\includegraphics[width=0.7\columnwidth]{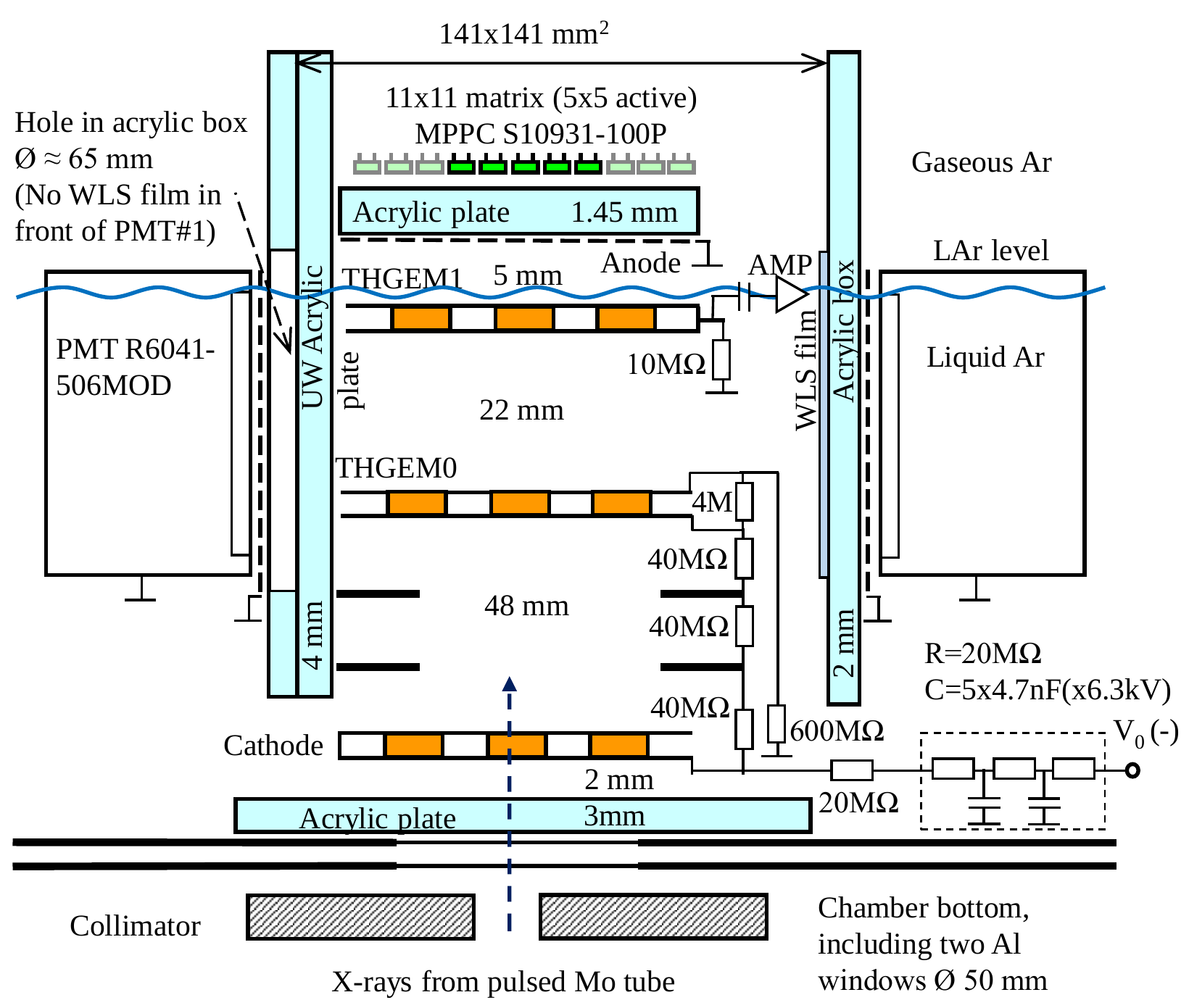}}
	\caption{Schematic view of the experimental setup.}
	\label{fig_setup}
\end{figure}

Fig.~\ref{fig_setup} shows the experimental setup. It was similar to those described in~\cite{Buzulutskov2018} and~\cite{Aalseth2020}, used for the studies of proportional electroluminescence in two-phase Ar detector and its SiPM-matrix readout respectively. The difference of the present setup as compared to the previous ones was that the cryogenic chamber was operated in a single-phase mode: it was filled with 3.5~liters of the liquid (pure argon or argon doped with methane) instead of 2.5~l when operated in the two-phase mode.
The detector was operated in equilibrium state at a saturated vapor pressure of 1.00~atm. For pure Ar this corresponds to a temperature of 87.4~K. For Ar+CH$_4$ mixture, the actual temperature depends on the CH$_4$ content.

To prepare the mixture, Ar and CH$_4$ were used with an initial purity of 99.9998\% and 99.95\%, respectively.
The measurements were conducted over several measurement sessions, each one day long. At the beginning of the session, the gas mixture from the stainless steel bottle was passed through an Oxisorb filter~\cite{spectron} for purification from electronegative impurities and then was liquefied into the cryogenic chamber. At the end of the session, the content of the cryogenic chamber was collected back into the bottle.

The first measurement sessions were done with pure Ar. 
To prepare a certain Ar+CH$_4$ mixture before the next session, the gas composition was changed by cooling the bottle with given mixture and adding the required amount of CH$_4$ or pure Ar.

The CH$_4$ content in the mixture thus prepared was measured using a Residual Gas Analyzer (RGA) Pfeiffer-Vacuum QME220. In this case, the bottle was connected to a baked high-vacuum ($4\times 10^{-9}$~mbar) test chamber equipped with RGA, where the CH$_{4}$ content in Ar was measured in a flow mode at a pressure reaching $10^{-4}$~mbar.
The relative error between the expected and measured content values was below 10\%.

The single-phase detector was in fact a liquid TPC composed of two gaps: that of the low electric field, 48~mm thick, and that of the high electric field, 22~mm thick. The presence of two gaps is explained by the fact that the detector was originally designed to operate in the two-phase mode, the second gap being operated in electroluminescent mode. 

To form these gaps, the electrodes made from THGEMs (Thick Gas Electron Multipliers,~\cite{Breskin2009}) were used: see Fig.~\ref{fig_setup}. 
The low-field gap was formed by a cathode electrode, field-shaping electrodes and THGEM0. These were biased through a resistive high-voltage divider placed within the liquid. The electron transmission efficiency, defined by the voltage applied across THGEM0 and its geometrical parameters, was calculated in~\cite{Bondar2019_field_sim}: it amounted to 61\%.
The high-field gap was formed by a THGEM0 and THGEM1 acted as an anode. All electrodes had the same active area of 10$\times$10~cm$^2$. 

The liquid level in the cryogenic chamber was between THGEM1 and acrylic plate in front of the SiPM matrix.
It was calculated from the amount of condensed Ar using CAD software and verified using THGEM1 as a capacitive liquid level meter. 

Three different spectral devices were used in the measurements. 
Four compact cryogenic PMTs R6041-506MOD~\cite{CryoPMT15} were located on the perimeter of the high-field gap and electrically insulated from it by an acrylic box.
Three of four PMTs were made sensitive to the VUV via WLS films (based on TPB in a polystyrene matrix~\cite{Gehman2013}) deposited on the inner box surface facing the EL gap, in front of these PMTs. Let us designate this spectral device as 3PMT+WLS.

The second spectral device was the bare PMT, without WLS. 
It was isolated from the EL gap by an UV-acrylic plate transparent above 300 nm. It is designated as 1PMT.

The third spectral device was a 5$\times$5 SiPM matrix. 
It was composed from SiPMs of MPPCs 13360-6050PE type~\cite{hamamatsu} with an active area of 6$\times$6~mm$^2$ each and channel pitch of 1~cm.

\begin{figure}[!htb]
	\center{\includegraphics[width=0.6\columnwidth]{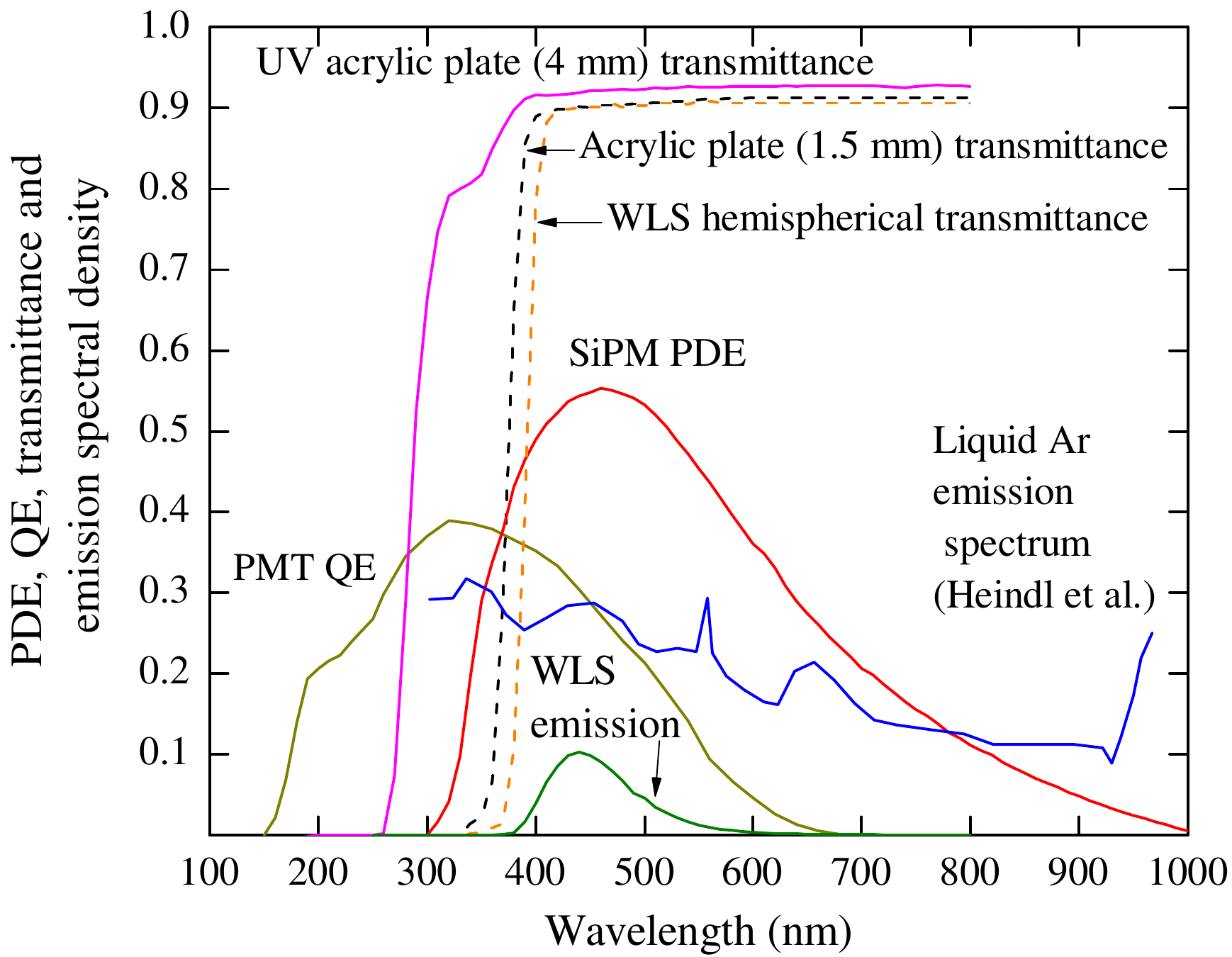}}
	\caption{Photon Detection Efficiency (PDE) of SiPM (MPPC 13360-6050PE~\cite{hamamatsu}) at overvoltage of 5.6~V obtained from~\cite{Otte2017} using the PDE voltage dependence, Quantum Efficiency (QE) of the PMT R6041-506MOD at 87~K obtained from~\cite{hamamatsu,Lyashenko2014} using a temperature dependence derived there, the transmittance of the ordinary and UV acrylic plate in front of the SiPM and 1PMT, respectively, and hemispherical transmittance of the WLS (TPB in polystyrene)~\cite{Francini2013}. Also shown are the scintillation spectrum of liquid Ar reported in~\cite{Heindl2011} and the emission spectrum of the WLS (TPB in polystyrene)~\cite{Gehman2013}.}
	\label{fig_stectra}
\end{figure}

Fig.~\ref{fig_stectra} shows the Photon Detection Efficiency (PDE) of SiPMs, Quantum Efficiency (QE) of PMTs, transmittance of the ordinary and UV acrylic plate in front of the SiPM and 1PMT, respectively, and hemispherical transmittance of the WLS~\cite{Francini2013}.
In addition, the scintillation spectrum of liquid Ar in the visible range~\cite{Heindl2011} and the emission spectrum of the WLS~\cite{Gehman2013} are presented.
Taking into account the transmission of the acrylic plate in front of the matrix, the SiPM-matrix sensitivity ranges from the near UV (360~nm) to the NIR (1000~nm).
The contribution of crosstalk from the VUV, re-emitted by WLS, was negligible, as shown by experiments under similar conditions without WLS.

The detector was irradiated from outside by X-rays from a pulsed X-ray tube with Mo anode, with the average deposited energy in liquid Ar of 25~keV~\cite{XRayYield16}.

The signals from the PMTs were amplified using fast 10-fold amplifiers CAEN N979 and then re-amplified with linear amplifiers with a shaping time of 200~ns. The signals from 3PMT+WLS were summed (using CAEN N625 unit). The signals from each SiPM were transmitted to amplifiers with a shaping time of 40~ns, via twisted pair wires. The charge signal from the THGEM1 anode was recorded using a calibrated chain of a preamplifier and shaping amplifier. All amplifiers were placed outside the two-phase detector.

The DAQ system included both a 4-channel oscilloscope LeCroy WR HRO 66Zi and a 64-channel Flash ADC CAEN V1740 (12~bits, 62.5~MHz): the signals were digitized and stored both in the oscilloscope and in a computer for further off-line analysis.

Other details of the experimental setup and measurement procedures can be found elsewhere~\cite{BondarINSTR2017EL,Buzulutskov2018,Aalseth2020}.

\section{Results}

\begin{figure}[!htb]
	\center{\includegraphics[width=0.7\columnwidth]{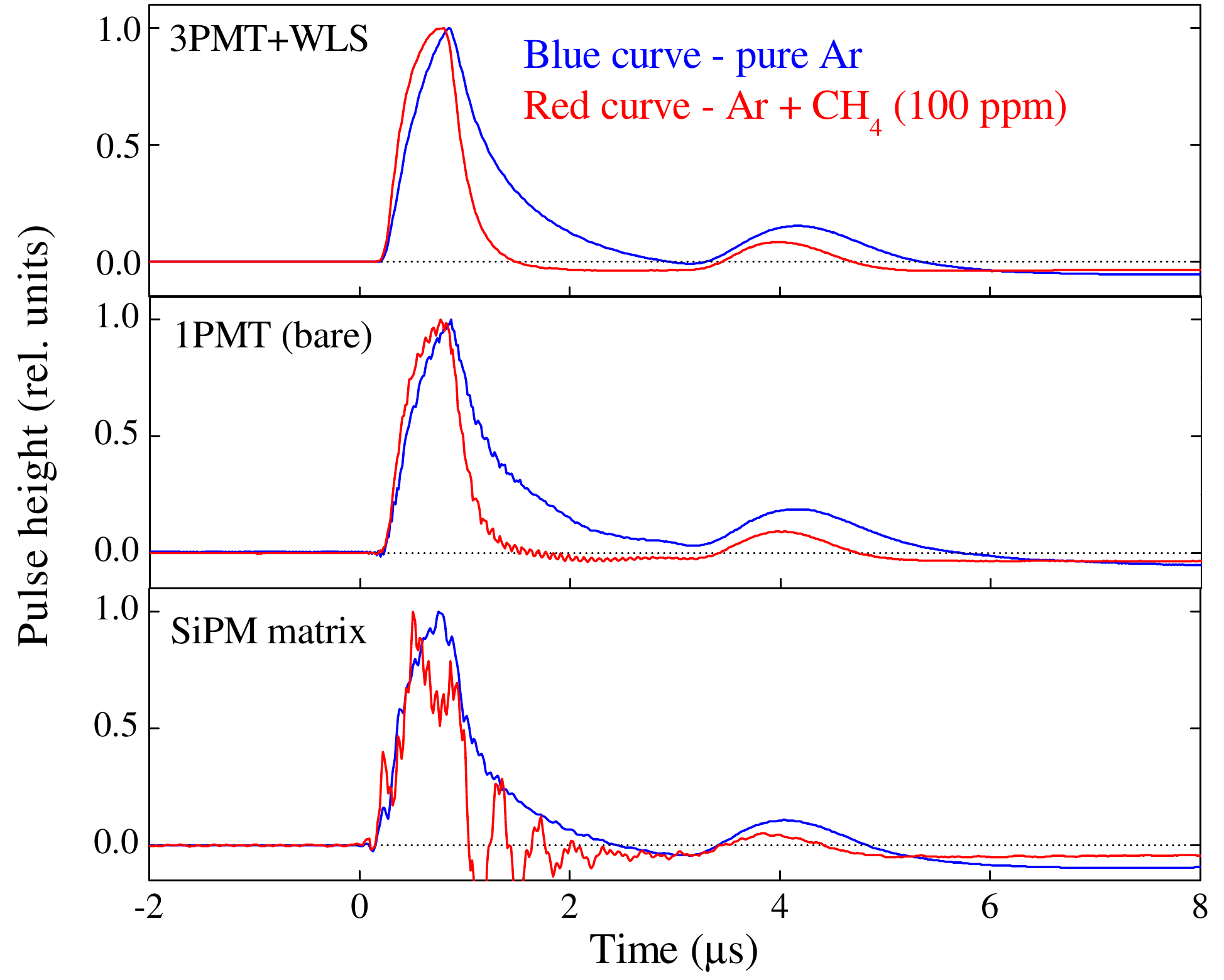}}
	\caption{Averaged pulse-shapes of the signals from PMTs with WLS (3PMT+WLS), bare PMT (1PMT) and SiPM matrix, induced by pulsed X-rays, obtained in liquid Ar and liquid mixture Ar+CH$_4$ (100 ppm) at zero electric field. All the pulse heights are normalized to the maximum.}
	\label{fig_pulse_shapes}
\end{figure}

We have observed the scintillation signals in pure liquid Ar and liquid Ar+CH$_4$ mixture, induced by pulsed X-rays, from all the three types of the spectral devices described above, i.e. including in the visible and NIR range. Their averaged pulse shapes obtained at zero electric field in the liquid are shown in Fig.~\ref{fig_pulse_shapes}. The double-peak pulse structure is due to characteristic pulse-shape of the X-ray tube itself~\cite{StudyInfraredScintillations2012}. The oscillations distorting the SiPM-matrix signal observed for the weak signal in Ar+CH$_4$ are due to the noise contribution. One can see that the signal in Ar+CH$_4$ is faster than in pure Ar.

Now let us estimate the (absolute) photoelectron yield ($Y_{PE}$) for the given spectral device (3PMT+WLS, 1PMT or SiPM matrix). It is defined as the ratio of the number of recorded photoelectrons ($N_{PE}$) and the primary ionization charge ($N_{i}$):
\begin{equation}
\label{eq_PE_yield} Y_{PE} = N_{PE} / N_{i} \, .
\end{equation}
The photoelectron number was integrated over the time of 10~$\mu$s from the beginning of the signal. 
In this paper, we focus on the study of the photoelectron yield at zero field, i.e. calculating the photoelectron number for signals obtained at zero electric field in the liquid.

The primary ionization $N_i$ charge was calculated from the dependence of the collected charge ($N_{coll}$) on the electric field $F$ in liquid Ar, accounting for the recombination effect using the well-known parametrization with the recombination coefficient $k_{rec}$ ~\cite{Barabash_1993,Aprile_2006,Chepel2013}:
\begin{equation}
\label{eq_charge_vs_field_full} N_{coll} = N_{i} \cdot \frac{T_e \cdot \exp(- K_{att} \cdot C \cdot X  )}{1 + k_{rec}/F} \, .
\end{equation}
Here in addition, both the electron transmission through the THGEM0 electrode ($T_e$=61\%) and the attachment of electrons drifting over the distance $X$ to electronegative impurities (with concentration $C$), using the attachment coefficient $K_{att}$, were taken into account. The latter was taken as~\cite{Aprile_2006}: 
\begin{equation}
\label{eq_Katt} K_{att} = 0.95 / F^{0.8} \, ,
\end{equation}
where $K_{att}$ is expressed in (ppm$\cdot$mm)$^{-1}$ and $F$ in kV/cm.

\begin{figure}[!htb]
	\center{\includegraphics[width=0.6\columnwidth]{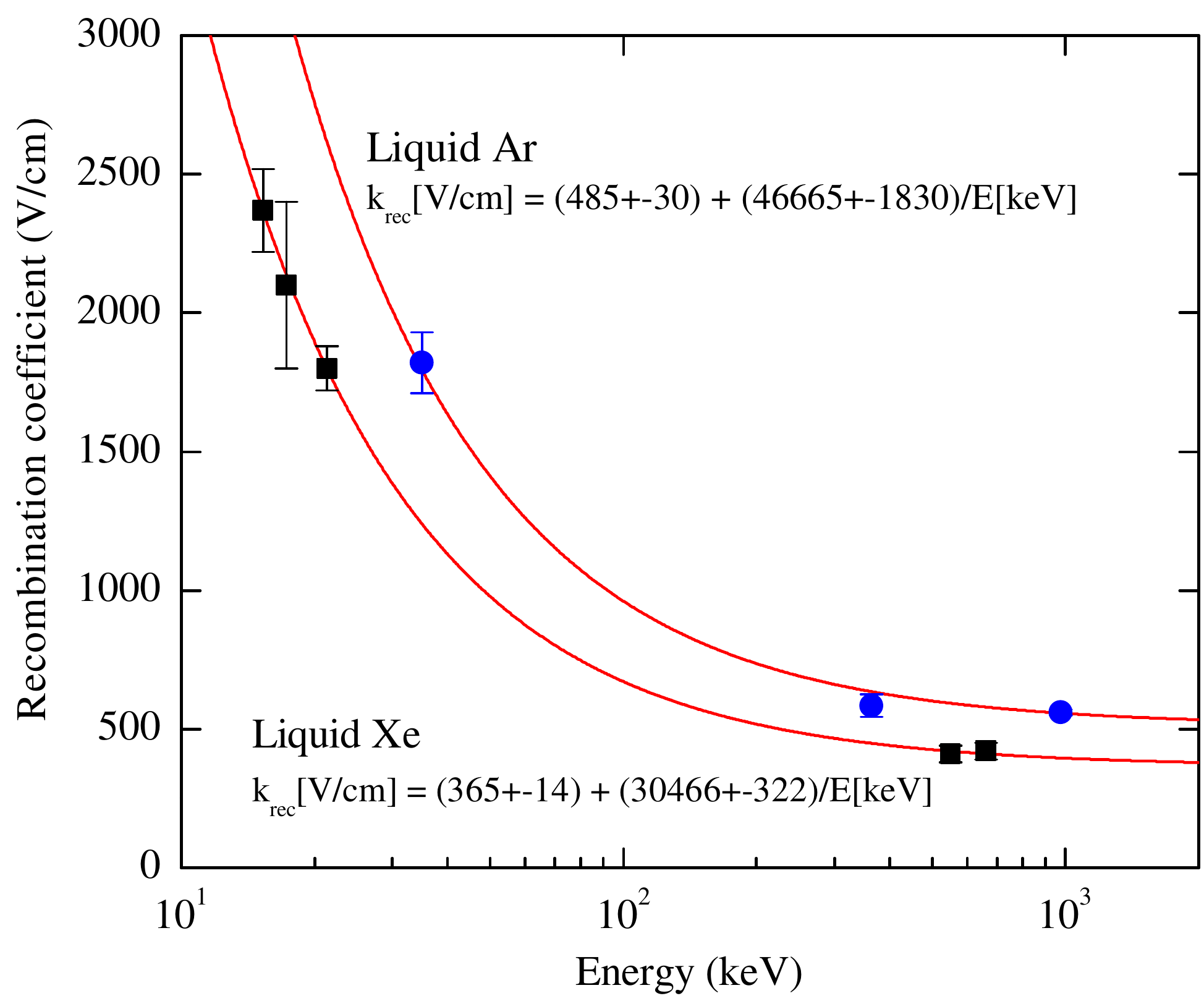}}
	\caption{Recombination coefficient in liquid Ar and Xe as a function of the X-ray or electron energy obtained from~\cite{XRayYield16}.}
	\label{fig_Kreco_vs_enegy}
\end{figure}

The recombination coefficient is a function of the energy: see 
Fig.~\ref{fig_Kreco_vs_enegy} showing its dependence on the X-ray or electron energy in liquid argon or xenon obtained from~\cite{XRayYield16}.
Since there are no experimental data in liquid Ar in the energy of interest of the present work (25 keV), the recombination coefficient was obtained by extrapolation, using the function similar to that used for liquid Xe where there are enough experimental data at the given energy.
The recombination coefficient for 25~keV X-rays in liquid Ar thus obtained amounted to 2.35~kV/cm.

\begin{figure}[!htb]
	\center{\includegraphics[width=0.6\columnwidth]{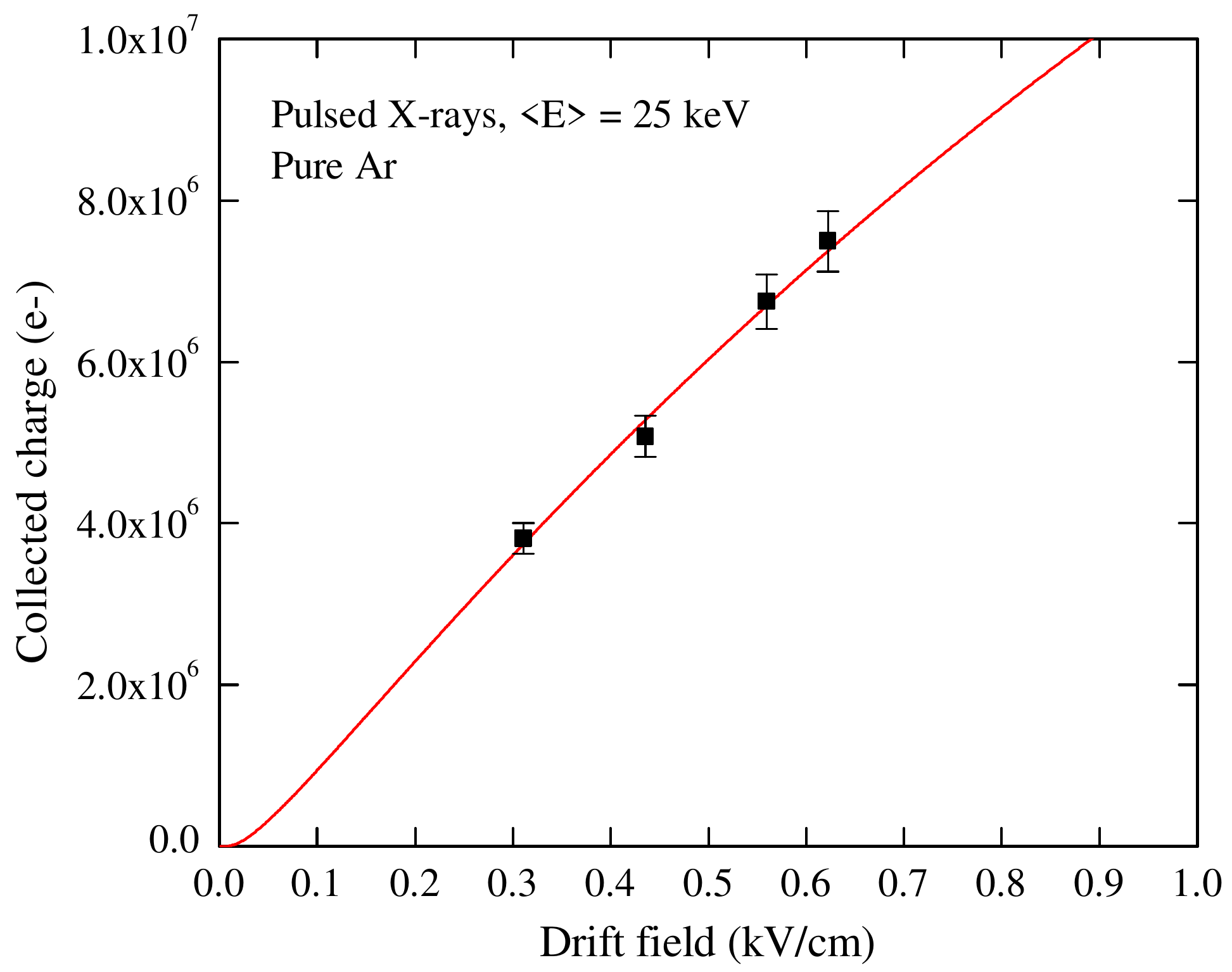}}
	\caption{Collected charge in pure liquid Ar as a function of the drift field. Red curve is the fit by~\eqref{eq_charge_vs_field_full}, where $N_{i}$ and $C$ are free parameters.}
	\label{fig_charge_vs_field_pureAr}
\end{figure}

The collected charge in pure Ar was fitted by~\eqref{eq_charge_vs_field_full}, where the primary ionization charge ($N_i$) and oxygen concentration ($C$) were used as free parameters: see Fig.~\ref{fig_charge_vs_field_pureAr}.
The obtained oxygen concentration amounted to about $2 \pm 0.8$~ppb, which corresponds to electron lifetime of $150 \pm 60 \ \mu s$ at a drift field of 200~V/cm, in particular used in the DarkSide experiment for dark matter search~\cite{Agnes2015}.
The obtained primary ionization charge amounted to $(65 \pm 5)\cdot 10^6$~e$^{-}$.

The primary ionization charge was measured this way only for pure Ar, since the attachment and recombination coefficients for the mixture Ar+CH$_4$ are unknown.
Since the relative position of the X-ray tube and the single-phase detector was fixed, we can assume that the primary ionization charge was the same for Ar+CH$_4$ mixtures, at least for small CH$_4$ contents ($<$10\%).
The systematic error for the primary ionization charge is estimated to be below 20\%.

\begin{figure}[!htb]
	\center{\includegraphics[width=0.7\columnwidth]{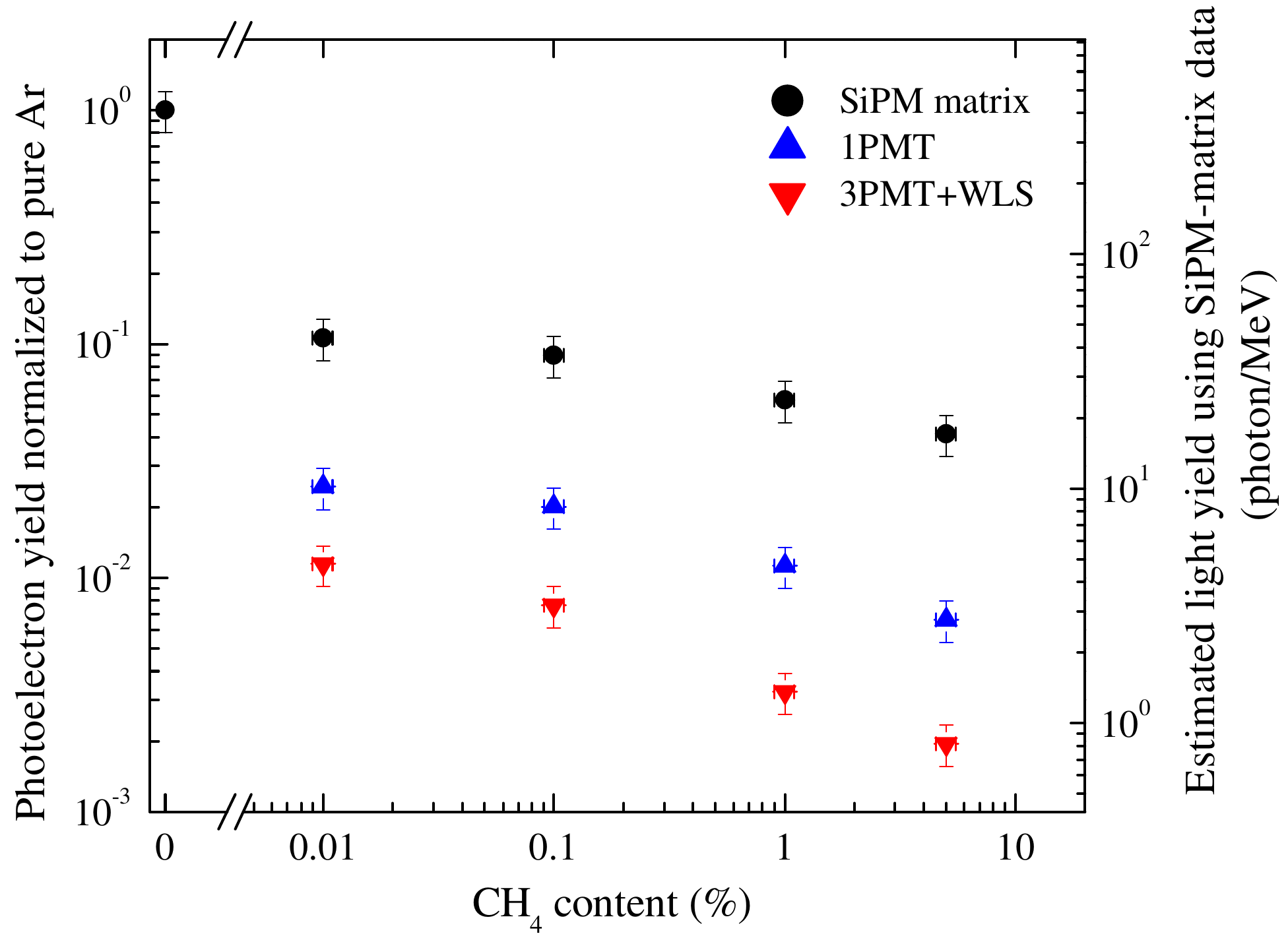}}
	\caption{Relative photoelectron yield of primary scintillations (induced by pulsed X-rays) in pure liquid Ar and its mixtures with CH$_4$ (left scale) as a function of the CH$_4$ content, for different readout spectral devices, namely for SiPM matrix, bare PMT (1PMT) and 3 PMT with WLS (3PMT+WLS). Also shown is the light yield estimated using SiPM-matrix data (right scale, relevant to the SiPM-matrix data points). Yields were obtained at zero electric field in the liquid.}
	\label{fig_rel_LY_TPB}
\end{figure}

Fig.~\ref{fig_rel_LY_TPB} shows the relative photoelectron yield, defined as photoelectron yield normalized to that of pure Ar, as a function of the CH$_4$ content.
In particular, the photoelectron yield in pure Ar at the SiPM matrix (25 central channels) amounted to $1.0 \cdot 10^{-6}$ PE/e$^{-}$.
In addition, the estimations for the light yield ($Y$) are shown in the same figure. 
Yields were obtained at zero electric field in the liquid.

The (absolute) light yield is defined as the ratio of the photoelectron yield ($Y_{PE}$) to the photon-to-photoelectron conversion efficiency ($PCE$): $Y = Y_{PE} / PCE$.
The conversion efficiency is defined as $PCE = \varepsilon < PDE >$. Here $\varepsilon$ is the photon collection efficiency, calculated using Monte Carlo simulation, and 
$< PDE >$ is the SiPM PDE averaged over the scintillation spectrum of liquid Ar taken from~\cite{Heindl2011} and  appropriately convolved with the acrylic transmittance spectrum (see Fig.~\ref{fig_stectra}). One can see from the figure that this experimentally measured emission spectrum is rather flat and resembles that of theoretical obtained for neutral bremsstrahlung effect in proportional electroluminescence~\cite{Buzulutskov2018}.

Accordingly, the light yield in pure Ar at zero field was estimated using the SiPM-matrix data and the photon emission spectrum from~\cite{Heindl2011}, in the wavelength range of 400-1000~nm: it amounted to $(10 \pm 1.4) \cdot 10^{-3}$~photon/e$^-$.
It is convenient to express the yield in the number of photons per MeV of deposited energy considering a W-value (energy needed to produce one ion pair) in liquid Ar of 23.6~eV~\cite{Miyajima1974}. 
The obtained value of $420 \pm 60$~photon/MeV is in good agreement with the result of $510 \pm 90$ photon/MeV reported for the same wavelength range elsewhere~\cite{Buzulutskov2011,StudyInfraredScintillations2012P2}.
The estimations of the light yield using other spectral devices, like 3PMT+WLS and bare PMT, had large systematic uncertainty since these in large extent were shaded by the field-shaping electrodes from the scintillation region (see Fig.~\ref{fig_setup}).

One can see from Fig.~\ref{fig_rel_LY_TPB} that adding even 100~ppm of CH$_4$ to pure Ar results in a considerable reduction of the photoelectron yield: by a factor of 10, 40 and 90 for SiPM-matrix, 1PMT and 3PMT+WLS readout, respectively.
As expected, the most significant reduction was observed for the 3PMT+WLS device, since it was sensitive to the excimer emission in the VUV which was completely quenched by the CH$_4$ molecules as discussed in Introduction. A smaller drop was observed for the bare PMT (1PMT), while the photoelectron yield for the SiPM matrix changed the least.

With further increase in methane content, from 100~ppm up to a maximum of 5\%, the yield decreased much more slowly: by a factor of 2.6, 3.7 and 5.9  for the SiPM matrix, 1PMT, and 3PMT+WLS readout, respectively. 

Using the SiPM-matrix data, we can estimate the light yield in liquid Ar+CH$_4$ mixtures, assuming  that the emission spectra in the visible and NIR range and  energy needed to produce one ion pair remained unchanged in presence of CH$_4$, compared to pure Ar. 
The latter statement is based on the fact that in gaseous pure Ar and pure CH$_4$ these energies are quite close~\cite{Tawara1987}.
In the absence of other data, such hypotheses seem reasonable.
Under such assumption, the light yield varied from 44~photon/MeV at 100~ppm of CH$_4$ to 17~photon/MeV at 5\% of CH$_4$. 

%At the CH$_4$ content of 5\%, the measured light yield amounted to $17 \pm 4$~photon/MeV.

One may conclude that the primary scintillations in liquid Ar doped with CH$_4$ indeed exist, in the visible and NIR range, though at rather weak level.
Since we know almost nothing about the nature of these non-VUV scintillations, we can only hypothesize that the NBrS mechanism might be responsible for these scintillations.

\section{Conclusions}

In this work, the primary scintillations have for the first time been observed in liquid argon doped with methane, in the visible and NIR range, induced by pulsed X-rays. Their relative light yields have been measured as a function of CH$_{4}$ content, the latter varying from 100~ppm to 5\%. 

The light yield for pure liquid Ar, defined at zero electric field, was estimated to be about 420~photon/MeV in the range of 400-1000~nm, which is in a good agreement with the previous result~\cite{Buzulutskov2011,StudyInfraredScintillations2012P2}.

However, when adding CH$_4$ to liquid Ar the estimated light yield at zero field was considerably reduced: down to about 44 and 17~photon/MeV at 100~ppm and 5\% of CH$_4$, respectively. 

These values seem to be too small for the effective detection of gamma-rays accompanying neutron capture on argon and hydrogen. As a result, a liquid mixture of argon and methane can hardly be considered as a working medium for neutron veto detectors.  
On the other, hand such a small light yield might be enough for hadron calorimetry, where the detector based on scintillating liquid mixture of argon and methane could be used, since methane allows to compensate the calorimeter and thus significantly increase its energy resolution.

Since we know almost nothing about the nature of these non-VUV scintillations, we can only hypothesize that the neutral bremsstrahlung (NBrS) mechanism, proposed elsewhere~\cite{Buzulutskov2018}, might be responsible for these scintillations.

Further studies of primary scintillations in liquid Ar mixtures with CH$_4$ are in the course in our laboratory.

%\section*{Acknowledgments}
\acknowledgments
This work was supported in part by Russian Science Foundation (project no. 20-12-00008).
It was done within the R\&D program for the DarkSide-20k experiment.

\bibliographystyle{JHEP}
\bibliography{mybibliography}

\end{document}